\documentclass[12pt]{article}
\usepackage{amsmath}
\usepackage{amsthm}
\usepackage{amssymb}
\usepackage{latexsym}
\usepackage{subeqnarray}
\usepackage{amsfonts}
\newcommand{\nc}{\newcommand*}

\nc{\reff}[1]{(\ref{#1})}
\nc{\nn}{\nonumber}
\nc{\ts}{\textstyle}
\nc{\ds}{\displaystyle}
\begin{document}

\centerline{\Large\bf On the spectrum of discrete Schr\"{o}dinger equation}
\smallskip

\centerline{\Large\bf  with one-dimensional perturbation\footnote{This article is extended variant of the
manuscript with the same title submitted to the Proceedings of the Days on Diffraction 2016}}
\bigskip

\centerline{\large\bf V. V. Borzov$^{\,a}$ E. V. Damaskinsky$^{\,b}$}
\smallskip

\begin{center}
{$^a$ Department of Mathematics, St.Petersburg State University of Telecommunications, \\ Prospekt
Bolshevikov 22-1, St.Petersburg, 193232, Russia; borzov.vadim@yandex.ru}

{$^b$ Mathematical Department, VI(IT), Russia, 191123, Zacharievskaya 22, St.Petersburg, Russia; evd@pdmi.ras.ru}
\end{center}

\centerline{\bf Abstract}
\begin{quote}
We consider the spectrum of the discrete Schr\"{o}dinger equation with one-dimensional
perturbation. We obtain the explicit form of scattering matrix and find the exact condition of
absence of singular part of the spectrum.
We calculated also the eigenvalue  that appears if this condition is not true.
In the last part of our paper we give few remarks on the case of
two-dimensional perturbations.
\end{quote}
\medskip

\section{Introduction}

The problem of a change in the spectrum of self-adjoint operator with one-dimensional
 perturbation in the well-known Friedrichs - Faddeev model \cite{1},\cite{2} on a finite segment $[a,b]$
 is well studied. Description of this model and references see in
the monograph \cite{3}. Let  $v(\lambda,\tau)=\beta u(\lambda)\overline{u(\tau)}$
is the kernel of the integral operator of the perturbation.
The function $v(\lambda,\tau)$  satisfies the H\"{o}lder condition with index $\alpha_0,$
and $u(a)=u(b)=0.$
It is well-known that in the case of one-dimensional perturbation  the absolutely continuous part
 of the spectrum not changed, and the singular part of the continuous spectrum for arbitrary
function $u(\lambda)$  does not only when $\alpha_0>\frac{1}{2}$.
It is also known that if the coupling constant $ \beta $ is small, then the discrete spectrum
of the perturbed operator from outside the segment $[a,b]$
is missing.  However, there is some positive critical value $ \beta_0 $ such that
if  $\vert \beta\vert> \beta_0 $, then there is
one simple eigenvalue $\lambda$ from
outside the segment $[a,b]$. In the general case, even for one-dimensional perturbations,
provided that $\alpha_0\leq\frac{1}{2}$ the singular spectrum can be  quite complicated
(see theorem 6.7.10 in \cite{3}).

In this paper we consider a special case of  one-dimensional perturbations
in  Friedrichs-Faddeev model on the finite segment $[-2,2]$. Namely the case when
the kernel of the integral operator of the perturbation takes the form
 $u(\lambda)=U_{k}(\frac{\lambda}{2})f(\lambda),$ ($k\in Z_{+}$)
where
$$
f(\lambda)=\frac{1}{\pi}\sqrt{1-\frac{\lambda^2}{4}}.
$$
and  $U_{n}$ are the Chebyshev polynomials of the 2-nd kind.
These polynomials  satisfy the following recurrence relations
\begin{equation}\label{triv03}
\lambda U_{n}(\ts\frac{\lambda}{2})=U_{n+1}(\frac{\ts\lambda}{2})+ U_{n-1}(\ts\frac{\lambda}{2}),\quad n\geq 1,
\qquad U_{0}(\ts\frac{\lambda}{2})=1,\qquad U_{1}(\ts\frac{\lambda}{2})=\lambda.
\end{equation}
In our case the function $u(\lambda)$ equals to zero at the ends of the segment and satisfies the
H\"{o}lder condition with index $\alpha_0=\frac{1}{2}.$

Taking into account possible applications of our results it is convenient to consider the
model as a discrete Schr\"{o}dinger operator with a local one-dimensional
 perturbation. This model is found in the study of some problems of atomic
physics (see \cite{7}), and also in the investigation of non-equivalent representations
algebra of generalized Chebyshev oscillator (see \cite{8}, \cite{8a}).

Let us turn to a precise formulation of the problem.
We give the definition of the considered Schr\"{o}dinger operator following \S\, 4.1
of the monograph \cite{4}.
Take as a starting point, the Jacobi matrix
$\mathcal{J}_{k}^{(\beta)}=\{\mathnormal{j}_{i,n}^{(\beta,k)}\}_{i,n=0}^{\infty},$
where
\begin{gather*}
\mathnormal{j}_{i,n}^{(\beta,k)}=
\left\{\begin{aligned}1,&\quad\text{if}\quad |i-n|=1 ,\\
\beta\delta_{nk},&\quad\text{if}\quad i=n ,\\
0,&\quad \text{if}\quad |i-n|>1,\end{aligned}\right.
\end{gather*}
 $\delta_{nk}$ is the Kronecker delta and $\beta$ is a real number.

Let $\mathcal{H}_{k}=\mathrm{L}_{2}(\mathcal{R};\mu_{k})$ be a separable Hilbert space
where  $\mu_{k}$ is a Borel probability measure,  such that
\begin{equation*}
 \nu_{n}=\int_{-\infty}^{+\infty}\lambda^{n} \mu_{k}(d\lambda)< \infty,\quad n\geq 0,
\end{equation*}
and $\nu_{0}=1.$
We consider $\mathcal{J}_{k}^{(\beta)}$ as the matrix of linear operator $H_{k}^{(\beta)}$
in $\mathcal{H}_{k}$
defined on the orthonormal basis  $\{\phi_{n}^{(k)}(\lambda)\}_{n=0}^{\infty}$ of the space  by relations
\begin{gather}\label{triv1}
H_{k}^{(\beta)}\phi_{n}^{(k)}\!=\!\phi_{n+1}^{(k)}\!+\!\phi_{n-1}^{(k)}\!+\!\beta\delta_{nk}\phi_{n}^{(k)},
\quad n=1,2,..,\\
H_{1}^{(\beta)}\phi_{0}^{(k)}=\beta\delta_{0k}\phi_{0}^{(k)}+ \phi_{1}^{(k)}.
\label{triv1`}
\end{gather}
The symmetric operator $H_{k}^{(\beta)}$ is defined on  the set of finite linear combinations
of basis vectors. The set is dense in the space $\mathcal{H}_{k}.$
 It is known \cite{4} that the deficiency indices of the operator $H_{k}^{(\beta)}$ are
equal to (0,0). Hence its closure is a self-adjoint operator with a simple spectrum in $\mathcal{H}_{k}$.
We denote this operator by the same symbol $H_{k}^{(\beta)}$. The operator $H_{k}^{(\beta)}$
defined by (\ref{triv1}), (\ref{triv1`}) is considered  discrete Schr\"{o}dinger operator
with a local one-dimensional  perturbation.

Further, denote by $\mathcal{J}_0$ the Jacobi matrix
$\mathcal{J}_0=\mathcal{J}_{k}^{(0)}$ and by $H_{0}$ the corresponding
self-adjoint operator $H_{0}=H_{k}^{(0)}$.

In the following we will assume that $\mathcal{H}_{k}$ is a space of
the spectral representation, i.e. the operator $H_{k}^{(\beta)}$
in the space $\mathcal{H}_{k}$  is the operator of multiplication by the variable $\lambda$\,.

It is known \cite{9} that the operator $H_{0}$ in spectral representation
defined on the Hilbert space $\mathcal{H}_{0}=\mathrm{L}_{2}(\mathcal{R};\mu_{0})$ with
\begin{equation}\label{4}
d\mu_{0}(\lambda)=\frac{1}{\pi}
\left\{\begin{aligned}\sqrt{1-\frac{\lambda^2}{4}}d\lambda,&\quad\text{if}\quad |\lambda|\leq 2,\\
0,\qquad&\quad \text{if}\quad |\lambda|>2,\end{aligned}\right.
\end{equation}
The Chebyshev polynomials of the 2-nd kind $\{U_{n}(\frac{\lambda}{2})\}_{n=0}^{\infty}$
form an orthonormal basis in the space $\mathcal{H}_{0}$.
"Perturbed" operator $H_{k}^{(\beta)}$ defined on the Hilbert space
$\mathcal{H}_{k},$ where the polynomials
$\{\phi_{n}^{(k)}(\lambda))\}_{n=0}^{\infty}$ form an orthonormal basis. These polynomials
satisfy the following recurrence relations
\begin{equation}\label{triv001}
\lambda\phi_{n}^{(k)}(\lambda)=\phi_{n+1}^{(k)}(\lambda)+\beta\delta_{nk}\phi_{n}^{(k)}(\lambda)+
\phi_{n-1}^{(k)}(\lambda),\quad n\geq 1,
\end{equation}
\begin{equation}\label{triv002}
\phi_{0}^{(k)}(\lambda)=1,\quad\phi_{1}^{(k)}(\lambda)=(\lambda-\beta\delta_{0k}).
\end{equation}

{\bf Our aim is to study the singular part of the spectrum of self-adjoint "perturbed"
Hamiltonian $H_{k}^{(\beta)}$,
($\beta\in \mathbb{R}$) for arbitrary  $k\in {Z_{+}}.$} This operator occurs  after
adding  one-dimensional perturbation to the "free" Hamiltonian  $H_{0}$
(corresponding to the case $\beta=0$).
We use a new representation of the resolvent $ R_{0}(z)=(H_{0}-z)^{-1}$ of self-adjoint
unperturbed  operator $H_{0}$ in $\mathcal{H}_{0}$ .
By (.,.) we denote the scalar product in the Hilbert space $\mathcal{H}_{0}$ .
 Computing the roots of the denominator of the resolvent
$$
D(z)= 1 + \beta(R_{0}(z)U_k,U_k)
$$
on the real axis, we obtain the following  results:

1. Exact condition of the absence of singular spectrum on the continuous spectrum (on the segment $[-2,2]$).

2. The critical value $\beta_{0}$ of the coupling constant $\beta$ equals to $\frac{1}{k+1}.$

3. The boundary points $\pm 2$ of the continuous spectrum are resonances.

4. The explicit dependence $\lambda=\lambda(\beta)$ of the coupling constant $\beta$
the  eigenvalue  $\lambda$ which  is  outside the segment $[-2,+2]$.

5. The explicit form of the scattering matrix.

Besides, we build the orthogonality measure $\mu_{k}$ of the polynomials
$\phi_{n}^{(k)}(t).$

\section{The orthogonality measure $d\mu_{k}$}
We consider Jacobi matrix $\mathcal{J}_{k}^{(\beta)}$  and the matrix
\begin{equation*}
B_k=\{\mathnormal{j}_{i,m}^{(k)}\}_{i,m=1}^{\infty},\quad \mathnormal{j}_{i,m}^{(k)}=\beta \delta_{i,k}\delta_{m,k}
\end{equation*}
for $\beta\in {\mathbb{R}}.$
Denote by $V_k$ the one-dimensional self-adjoint operator corresponding to the matrix $B_k$
in a Hilbert space $\mathcal{H}_{0}.$ We consider the "perturbed" operator $H_{k}^{(\beta)},$
corresponding to the Jacobi matrix $\mathcal{J}_{k}^{(\beta)}$ in $\mathcal{H}_{0},$ as a sum of
$H_{0}$ and $V_k$:
\begin{equation*}
H_{k}^{(\beta)}=H_{0}+V_k,\quad  V_k=\beta (.,U_{k})U_{k},\quad\quad
\Vert U_{k}\Vert=1,\quad \beta=\overline{\beta}.
\end{equation*}
We will compute the  density $\mu_k^{\prime}(\lambda)$  of a measure  $\mu_{k}$.
Let $E_{0}(\lambda)\!=\!E_{0}(-\infty,\lambda)$ and
$E_{k}^{(\beta)}(\lambda)\!=\!E_{k}^{(\beta)}(-\infty,\lambda)$ ---
generating functions of the spectral measures $E_0$ and $E_{k}^{(\beta)}$ of self-adjoint
operators $H_0$ and $H_{k}^{(\beta)}.$
Let's denote by $\it{L_k},\it{L}_{k}^{(\beta)}$ --- linear spans of the sets
$\{E_{0}(X)U_k\}$ and $\{E_{k}^{(\beta)}(X)U_k\}$, respectively.
Then restrictions $\tilde{H}_{0},\,\, \tilde{H}_{k}^{(\beta)}\!,$ of the operators $H_{0}$ and
$H_{k}^{(\beta)}$ to the subspace $\it{L_0} =\it{L}_{k}^{(\beta)}$ of Hilbert space
$\mathcal{H}_{0}$  have a simple spectrum.
Therefore, they are multiplication operators on $\lambda$ in
$\mathrm{L}_{2}(\textbf{R};\varrho_{0}^{(k)})$ and $\mathrm{L}_{2}(\textbf{R};\varrho_{k}),$
where the measure $\varrho_{0}^{(k)}$ (we denote $\varrho_{0}^{(0)}$ by $\varrho_{0}$) and $\varrho_{k}$
are defined by  equalities
\begin{equation*}
\varrho_{0}^{(k)}(.)\!=\! (E_{0}(.)U_k,U_k),\quad\!
\varrho_{k}(.)\!=\!( E_{k}^{(\beta)}(.)U_k,U_k).
\end{equation*}
We introduce also the measure $\varrho_{k,0}$
\begin{eqnarray}
\varrho_{k,0}(.)= ( E_{k}^{(\beta)}(.)U_0,U_0).
\end{eqnarray}
It is known that
\begin{equation*}
\varrho_{0}(.)=\mu_0(.),\quad \varrho_{k,0}(.)=\mu_{k}(.).
\end{equation*}
Then we have
\begin{align}\label{triv104}
{\varrho_{0}^{(k)}}^{\prime}(\lambda)={\vert U_{k}(\ts\frac{\lambda}{2})\vert}^2\mu_{0}^{\prime}(\lambda), \\[4pt]
{\varrho_{k}}^{\prime}(\lambda)={\vert U_{k}(\ts\frac{\lambda}{2})\vert}^2\mu_{k}^{\prime}(\lambda),
\label{triv105}
\end{align}
where the measure $\mu_0$ defined by \reff{4}.
We consider the resolvent
$R_k(z)=(H_{k}^{(\beta)}-z)^{-1}$ of self-adjoint operator
$H_{k}^{(\beta)}$ in $\mathcal{H}_{0}$, and denote by $D_k(z)$ the following expression
\begin{equation*}
D_k(z)=1+\beta (R_0(z)U_{k},U_{k}).
\end{equation*}
From (\cite{3})
we know that $D_k(\lambda+i 0)\neq 0$ for almost all $\lambda$ and
\begin{equation}\label{triv77`}
{\rho_0^{(k)}}^{\prime}(\lambda)={\vert D_k(\lambda+i0)\vert}^2 \rho_k^{\prime}(\lambda).
\end{equation}
Then we get from (\ref{triv104}),(\ref{triv105}) , (\ref{triv77`}) for almost all $\lambda $
\begin{equation}\label{triv10}
\mu_k^{\prime}(\lambda)=\frac{\mu_0^{\prime}(\lambda)}{{\vert D_k(\lambda+i 0)\vert}^2}.
\end{equation}
From (\ref{triv77`}) and (\ref{triv10}) it follows  that  absolutely continuous parts
of the measures $\varrho_{0,a}^{(k)}$ and $\varrho_{k,a}$ are equivalent, i.e.
$\sigma_{a}(H_{k}^{(\beta)})=\sigma_{0,a}=[-2,2].$

By the Privalov theorem \cite{5},
using the well-known expression
for the resolvent of the operator $H_{0}$ via the Cauchy-Stieltjes integral with respect to
the spectral measure, we have
\begin{gather*}
(R_0(\lambda\pm i 0)U_k,U_k)=\pm\pi i
{\rho_0^{(k)}}^{\prime}(\lambda)+I^{(k)}(\lambda)\\
I^{(k)}(\lambda)= v.p.\int_{-\infty}^{+\infty}(t-\lambda)^{-1}
{\rho_0^{(k)}}^{\prime}(t)\,dt,
\end{gather*}
 We get from (\ref{triv104}) that
\begin{equation}\label{triv505}
I^{(k)}(\lambda)= v.p.\int_{-2}^{+2}(t-\lambda)^{-1}
\left(U_{k}(\ts\frac{t}{2})\right)^2\mu_{0}^{\prime}(t)\,dt,
\end{equation}

Using recurrence relations \reff{triv03}, for the Chebyshev polynomials one can calculate
the integral $I^{(k)}(\lambda)$\footnote{See Appendices 1\& 2 for the proof}
\begin{gather}
I^{(k)}(\lambda)=
\left\{\begin{aligned}
b_{k}(\lambda)U_{k}(\ts{\frac{\lambda}{2}}), & \quad\text{if}\quad |\lambda|\leq 2,\\
(-1)^{k}a^{k+1}(\lambda)U_{k}(\ts{\frac{\lambda}{2}}),& \quad\text{if}\quad |\lambda|>2,\end{aligned}\right.
\label{triv30}
\end{gather}
where
\begin{equation}
b_{k}(\lambda)=-\frac{\lambda}{2}\,U_{k}(\ts\frac{\lambda}{2})+U_{k-1}(\ts\frac{\lambda}{2}),
\end{equation}
and
\begin{equation}\label{triv05}
a(\lambda)=-\frac{2}{\lambda}\,\,\frac{1}{1+\sqrt{1-\frac{4}{\lambda^2}}}\,.
\end{equation}
We have for $|\lambda|\leq 2  $
\begin{equation*}
D_k(\lambda+i 0)=1+\beta\left(\pm\pi i\mu_{0}^{\prime}(\lambda)U_{k}^{2}(\ts\frac{\lambda}{2})+
\widetilde{L}_k(\lambda)\right),
\end{equation*}
\begin{equation}
\widetilde{L}_k(\lambda)=-\frac{\lambda}{2}\,U_{k}^{2}(\ts\frac{\lambda}{2})+
U_{k}(\ts\frac{\lambda}{2})\,U_{k-1}(\ts\frac{\lambda}{2}).
\label{triv130}
\end{equation}
In the case when $|\lambda|>2$ we have
\begin{equation*}
D_k(\lambda+i 0)\!=\!1+\beta(-1)^{k}a^{k+1}(\lambda)U_{k}(\ts\frac{\lambda}{2}).
\end{equation*}
Using the identity\footnote{See Appendix 3 for proof}
\begin{equation}
U_{k}^{2}\big(\ts{\frac{\lambda}{2}}\big)+U_{k-1}^{2}\big(\frac{\lambda}{2}\big)-
\lambda U_{k-1}\big(\frac{\lambda}{2}\big)U_{k}\big(\frac{\lambda}{2}\big)=1,
\label{triv405}
\end{equation}
we get
\begin{equation}\label{triv45}
{\vert D_k(\lambda+i0)\vert}^2\!=\!
\left\{\!\begin{aligned}
1\!+\!\beta(\beta\! -\! \lambda)A_{k}\!+\!
2\beta B_k,\quad &|\lambda|\leq 2,\\
(1+(-1)^{k}\beta a^{k+1}(\lambda)A_k,\quad & |\lambda|>2.
\end{aligned}\right.
\end{equation}
where
$$
A_k=U_{k}^{2}(\ts\frac{\lambda}{2}), \quad B_k=U_{k}(\frac{\lambda}{2})U_{k-1}(\ts\frac{\lambda}{2}).
$$
Finally, from \ref{4}, (\ref{triv10}) and (\ref{triv45})
it follows that
\begin{equation*}
\mu_{k}^{\prime}(\lambda)=
\frac{1}{\pi}
\left\{\begin{aligned}
\frac{\sqrt{1-\frac{\lambda^2}{4}}}
{1+\beta(\beta - \lambda)A_{k}+2\beta B_k},&\quad\text{if}\quad |\lambda|\leq 2,\\
0,\qquad\qquad\qquad &\quad\text{if}\quad |\lambda|>2.
\end{aligned}\right.
\end{equation*}

\section{Scattering Matrix for ($H_{0},H_{k}^{(\beta)}$)}

Now we turn to the scattering matrix. According to theorem 6.7.3 (\cite{3})
the scattering  matrix $S(\lambda)$ for a pair of self-adjoint operators $H_{0}$ and $H_{k}^{(\beta)}$
is calculated for almost all $\lambda\in\widehat{\sigma}_0$ by the following formula
\begin{equation}\label{triv14`}
S^{(k)}(\lambda)=I(\lambda)- 2\pi\beta i D_{k}^{-1}(\lambda+i 0)
\langle .,\widetilde{U}_{k} (\frac{\lambda}{2})\rangle_{h(\lambda)}
\widetilde{U}_{k} (\frac{\lambda}{2}),
\end{equation}
Here $\widehat{\sigma}_0=[-2,2]$ is the core of the spectrum of the operator $H_{0}$
(the minimal Borel support of the spectral measures $E_0$).
By $\langle ., .\rangle_{h(\lambda)}$ we denote
the scalar product in "infinitesimal subspace" $h(\lambda)$ of a direct integral which corresponds
to the absolutely continuous part of the operator $H_{0}$. Namely, we
denote by $H_{0}^{(a)}$ the restriction of the operator $H_{0}$ on
the absolutely continuous subspace $\mathcal{H}_{0}^{(a)}$ of the operator $H_{0}$ and  consider
the decomposition of the subspace $\mathcal{H}_{0}^{(a)}$ into a direct integral (see \cite{3} for details)
\begin{equation}
\mathcal{H}_{0}^{(a)}\longrightarrow \int_{\widehat{\sigma}_0}^{}\oplus h(\lambda)\,d\lambda.
\label{triv15}
\end{equation}

The element $\widetilde{U}_{k}(.)$ in the formula (\ref{triv14`}) is the representative element of
$U_k$ in the decomposition (\ref{triv15}). In our case
\begin{equation}
\widetilde{U}_{k}(\ts\frac{\lambda}{2})=\ds\frac{1}{\sqrt{\pi}}\,
\sqrt[4]{1-\ds\frac{t^2}{4}}\,U_{k}(\ts\frac{\lambda}{2}) \quad\text{if}\quad
|\lambda|\leq 2.
\label{triv16`}
\end{equation}
Substituting (\ref{triv16`}) in (\ref{triv14`}), we get for almost all $\lambda\in [-2,2]$
\begin{equation*}
S^{(k)}= I-\widetilde{I}, \quad |\lambda|\leq 2,
\end{equation*}
where
$$
\widetilde{I}=
\frac{2\beta\sqrt{\pi}\,i\,\sqrt[4]{1-\frac{{\lambda}^2}{4}}\,U_{k}(\frac{\lambda}{2})}
{1+\beta(i\,\sqrt{1-\frac{\lambda^2}{4}}\,
U_{k}^{2}(\frac{\lambda}{2})+\widetilde{L}_k(\lambda))}\langle .,\widetilde{U}_{k}\rangle_{h(\lambda)}
$$
and $\widetilde{L}_k$ is defined by (\ref{triv130}).

\section{Point spectrum of the operator $H_{k}^{(\beta)}$}

According to  6.7.6 \cite{3}, if the point $ \lambda_0 $ belongs to the point spectrum
$\sigma(H_{k}^{(\beta)})$ on the segment $[-2,2]$, then it is a solution of the equation
\begin{equation}
{\vert D_k(\lambda+i0)\vert}^2 = 0.
\label{triv49}
\end{equation}

According to (\ref{triv45}), we rewrite this equation in the form
$$
1+\beta(\beta - \lambda)U_{k}^{2}(\ts\frac{\lambda}{2})+
2\beta U_{k}(\ts\frac{\lambda}{2})U_{k-1}(\ts\frac{\lambda}{2})=0.
$$
This is a quadratic equation with respect to the variable $\beta$:
\begin{equation}
U_{k}^{2}(\ts\frac{\lambda}{2}) \beta^{2}-U_{k} (\ts\frac{\lambda}{2})
\left(\lambda U_{k}(\ts\frac{\lambda}{2})-2U_{k-1}(\ts\frac{\lambda}{2})\right)\beta +1=0.
\label{triv59}
\end{equation}
For any $\lambda$ satisfying the inequality $U_{k}(\frac{\lambda}{2})\neq 0$
it is easy to obtain the solution of the quadratic equation
\begin{equation*}
\beta_{\pm}(\lambda)=\frac{\lambda U_{k}(\frac{\lambda}{2})-
2U_{k-1}(\frac{\lambda}{2})}{2U_{k}(\frac{\lambda}{2})}
\pm \sqrt{\frac{\lambda^2}{4}-1}.
\end{equation*}
Since $\beta\in \mathbb{R}$, from (\ref{triv59}) it is clear that if $
\vert \lambda \vert < 2 $ solutions of equation (\ref{triv49}) does not exist.
Hence, we proved the absence of the point spectrum of the operator $H_{k}^{(\beta)}$ in the interval (-2,2).

Consider the boundary points $\pm 2$. Those numbers are solutions of the equation (\ref{triv59}).
More precisely, since
\begin{equation}\label{triv008}
U_{k}(1)=k+1, \quad       U_{k}(-1)=(-1)^{k}(k+1),
\end{equation}
then
\begin{equation*}
\beta_{\pm}(2)=\frac{1}{k+1}, \quad   \beta_{\pm}(-2) =-\frac{1}{k+1}.
\end{equation*}
Now we will show that  points $\pm 2$ are not eigenvalues of the operator $H_{k}^{(\pm \frac{1}{k+1})}.$

Using recurrence relations  (\ref{triv001}), (\ref{triv002}) it is easy to show that
the polynomials $\phi_{n}^{(k)}$
are calculated from the following formulas\footnote{See Appendix 4 for the proof}
\begin{gather}
\phi_{0}^{(k)}=U_{0},\, \phi_{1}^{(k)}=U_{1}, ...,\phi_{k}^{(k)}=U_{k},\quad\nonumber\\[3pt]
\phi_{k+s}^{(k)}=U_{k+s}-\beta U_{k-1+s}-\beta U_{k-3+s}-
\ldots -\beta U_{k+1-s},\qquad 1\leq s\leq k+1,\label{t25}\\[3pt]
\phi_{n}^{(k)}=U_{n}-\beta U_{n-1}-\beta U_{n-3}-
\ldots -\beta U_{n-(2k+1)},\qquad n\geq 2k+2.\nonumber
\end{gather}

From  \reff{triv008} and
\reff{t25} it follows that
\begin{gather*}
\sum_{n=0}^{\infty}{\vert \phi_{n}^{(k)}(\pm 2)\vert}^2\geq
\sum_{n=2k+2}^{\infty}{\vert \phi_{n}^{(k)}(\pm 2)\vert}^2 =
\infty,
\end{gather*}
 since for $n\geq 2k+2$ we have
\begin{equation*}
\left|\phi_{n}^{(k)}(\pm 2)\right|^2=(k+1)^2.
\end{equation*}
Thus the boundary points $\pm 2$ of the continuous spectrum are not
 eigenvalues of the operator $H_{k}^{(\pm 1/{k+1})}.$

Now we have to calculate the eigenvalues $\lambda(\beta)$ of the operator $H_{k}^{(\beta)}$
for $\vert\beta\vert > \frac{1}{k+1},$ i.e. outside the segment  $[-2,+2]$.
For this we need to find the solution of equation (\ref{triv49}) outside of segment $[-2,+2]$.
Taking into account (\ref{triv45}), the equation (\ref{triv49}) can be written outside of segment
 $[-2,+2]$ in the following way
\begin{equation}\label{triv109}
1+(-1)^{k}\beta a^{k+1}(\lambda)U_{k}(\ts\frac{\lambda}{2})=0.
\end{equation}
As can be seen from (\ref{triv05}), the function  $a^{k+1}(\lambda)\neq 0$  for $|\lambda|>2$.
Then we can find a solution of the equation (\ref{triv109})
\begin{equation*}
\beta=(-1)^{k+1}\,\frac{-\lambda\left(1+\sqrt{1-\frac{4}{\lambda^2}}\,\right)^{k+1}}{2}
\,\,\frac{(-\lambda/2)^k}{U_{k}(\lambda/2)}.
\end{equation*}
It is easy to see that
\begin{equation*}
\beta(2)=\frac{1}{k+1}, \quad   \beta(-2) =-\frac{1}{k+1},
\end{equation*}
and the function $ \vert\beta(\lambda)\vert $ is monotonically increasing
from $\frac{1}{k+1}$ to $\infty$ if $|\lambda|$ increases monotonically from $2$ to $\infty$.
Therefore the function $\beta(\lambda)$ has a unique inverse function
$\lambda=\lambda(\beta)$ and $\vert\beta(\lambda)\vert >\frac{1}{k+1}$ as $|\lambda|>2.$
The function $\lambda=\lambda(\beta)$ is the sought eigenvalue.

As shown above when $\vert\beta\vert\leq \frac{1}{k+1}$,
the operator $H_{k}^{(\beta)}$  has no point spectrum  on the segment $[-2,+2]$.
Besides, for any $\beta$ such that $\vert\beta\vert > \frac{1}{k+1}$ there is some eigenvalue
$ \lambda(\beta)$ of the operator $H_{k}^{(\beta)}$ on the interval $(2,+\infty)$
(if $\beta > \frac{1}{k+1}$) or  on the interval $(-\infty,-2)$ (if $\beta < - \frac{1}{k+1}$).

This means that the boundary points $\pm 2$ of the continuous spectrum are resonances of
operators $H_{k}^{(\pm\frac{1}{k+1})},$  respectively.

\section{Two-dimensional perturbation:
Case $\vec{k}=(1,2)$}
Now we consider a two-dimensional perturbation of the operator $H_{0},$ which
is determined by the Jacobi matrix
$$
B_{1,2}=\{\mathnormal{j}_{i,m}^{(1,2)}\}_{i,m=1}^{\infty},\,\,
\mathnormal{j}_{i,m}^{(1,2)}\!=\beta_{1}\delta_{i 1}\delta_{m 1}+
\beta_{2}\delta_{i 2}\delta_{m 2}
$$
for $\beta_1,\beta_{2}\in {\textbf{R}}.$
Denote by $V_{1,2}$ two-dimensional self-adjoint operator, corresponding to the matrix $B_{1,2}$
in the Hilbert space $\mathcal{H}_{0}.$
In addition, we consider the Jacobi matrix
\begin{equation*}
\mathcal{J}_{1,2}^{(\beta_1,\beta_2)}=\{\mathnormal{j}_{i,m}^{(\beta_1,\beta_2)}\}_{i,m=1}^{\infty},
\end{equation*}
where
\begin{gather*}
\mathnormal{j}_{i,m}^{(\beta_1,\beta_2)}\!=\!
\left\{\begin{aligned}1,&\quad\text{if}\quad |i-m|=1 ,\\
\beta_{1}\delta_{i1}\delta_{m1}+\beta_{2}\delta_{i2}\delta_{m2},
&\quad\text{if}\quad |i-m|\neq1 .\end{aligned}\right.
\end{gather*}

Denote by $\Phi_{(1,2)}=\{\phi_{n}^{(1,2)}(t)\}_{n=0}^{\infty}$  the set of Jacobi polynomials
related with  the matrix $\mathcal{J}_{1,2}^{(\beta_1,\beta_2)}.$ The polynomials $ \Phi_{(1,2)}$ are defined
the following recurrence relations
\begin{gather*}
t\phi_{n}^{(1,2)}(t)=\phi_{n+1}^{(1,2)}(t)+a_{n}\phi_{n}^{(1,2)}(t)+b_{n-1}
\phi_{n-1}^{(1,2)}(t),\\
\phi_{0}^{(1,2)}=1,\,n\geq 0,\\
b_n=1-\delta_{n,0},\, a_0=\beta_1,\, a_1=\beta_2,\, a_n=0,\, n\geq 2.
\end{gather*}
The Jacobi matrix $\mathcal{J}_{1,2}^{(\beta_1,\beta_2)}$
corresponds to the operator $H_{1,2}^{(\beta_1,\beta_2)}$
\begin{gather*}
H_{1,2}^{(\beta_1,\beta_2)}=H_{0}+V_{1,2},\\
V_{1,2}=\beta_1 (.,U_0)U_0+\beta_2 (.,U_1)U_1.
\end{gather*}
The polynomials $\Phi_{(1,2)}$  form an orthonormal in
$\mathcal{H}_{12}=\mathrm{L}_{2}(\textbf{R};d\mu_{1,2}(t)).$
Just as it was done previously, we can calculate the measure $\mu_{1,2}$
\begin{gather*}
\mu_{1,2}^{\prime}(t)=
\left\{\begin{aligned}\frac{\frac{1}{\pi}\sqrt{1-\frac{t^2}{4}}}{Q},&\quad |t|\leq 2,\\
0,\qquad &\quad |t|>2,\end{aligned}\right.
\end{gather*}
\begin{equation*}
Q=\beta_1^{2}+(1-\beta_{1}\beta_2)^2+
t\left(2\beta_{2}-\beta_1(1+\beta_{1}\beta_2+2\beta_2^2)\right)+
   t^{2}\beta_{2}(\beta_{2}+2\beta_1) -t^{3}\beta_2.
\end{equation*}

{\bf Remarks }

1. Note that if $\beta_2=0$, we have $\mu_{1,2}^{\prime}(t)=\mu_{1}^{\prime}(t)$ and
if $\beta_1=0$, we have $\mu_{1,2}^{\prime}(t)=\mu_{2}^{\prime}(t).$

2. In contrast to the case of one-dimensional perturbations,  for two-dimensional case
we can't say that the orthogonality measure  $\mu_{1,2}$ for
polynomials $\Phi_{(1,2)}(t)=\{\phi_{n}^{(1,2)}(t))\}_{n=0}^{\infty}$ in the Hilbert space
$\mathcal{H}_{12}=\mathrm{L}_{2}(\textbf{R}; d\mu_{1,2}(t))$
 coincides with   absolutely continuous part of the measure $\mu_{1,2}$
for some real $\beta_1\neq 0,\beta_2\neq 0$.

3. However,  when
$\vert\beta_1\vert\leq 1,\quad\beta_2=0$ or
$\vert\beta_2\vert\leq\frac{1}{2},\quad\beta_1=0$, and $\beta_1,\beta_2\in {\textbf{R}},$
the orthogonality measure
$\mu_{1,2}$ of the polynomials $\Phi_{(1,2)}(t)=\{\phi_{n}^{(1,2)}(t))\}_{n=0}^{\infty}$
in $\mathcal{H}_{k}$  coincides with  absolutely continuous
part of the measures $\mu_{1,2}$.

\smallskip

\section*{Acknowledgement}
 EVD grateful to RFBR for financial support under the grant 15-01-03148.

\section*{Appendix 1. The proof  of  equation \reff{triv30} when $\vert\lambda\vert >2$}
Here we prove the relation
\begin{equation}\label{n01}
I^{(k)}(\lambda)=\int_{-2}^{+2}\frac{1}{(t-\lambda)}
\left(U_{k}(\ts\frac{t}{2})\right)^2\mu_{0}^{\prime}(t)\,dt
=(-1)^{k}a^{k+1}(\lambda)U_{k}(\ts{\frac{\lambda}{2}}), \quad\text{if}\quad |\lambda|>2.
\end{equation}
For this aim we on the first step proved the additional relation
\begin{equation}\label{n02}
I^{(k)}(\lambda)=\int_{-2}^{+2}(\frac{1}{(t-\lambda)}
\left(U_{k}(\ts\frac{t}{2})\right)^2\mu_{0}^{\prime}(t)\,dt
=a\,\,\frac{1-a^{2(k+1)}}{1-a^2}, \quad\text{if}\quad |\lambda|>2,
\end{equation}
where $a=a(\lambda)$ defined by \reff{triv05} as
\begin{equation}\label{n03}
a(\lambda)=-\frac{2}{\lambda}\,\,\frac{1}{1+\sqrt{1-\frac{4}{\lambda^2}}}\,,
\end{equation}
from which it follows that
\begin{equation}\label{n04}
\lambda=-\frac{a^{2}+1}{a}.
\end{equation}

On the second step we demonstrate that right hand sides of the relations \reff{n01} and \reff{n02} are equal, that proves the formula \reff{n01}.

We will  prove formula \reff{n02} by induction. First we will check the validity of \reff{n02} for $k=0$  and $k=1$.
After the change of variable $t=2\cos x$ in the integral $I^{(0)}(\lambda)$ we obtain
\begin{equation*}
I^{(0)}(\lambda)=\int_{-2}^{+2}\frac{1}{(t-\lambda)} \mu_{0}^{\prime}(t)\,dt=
\frac{2}{\pi}\int_{0}^{\pi}\frac{\sin^{2}x}{-\lambda+2\cos x}\,dx.
\end{equation*}
According to eq. (3.644.4) in \cite{10} at $p=-\lambda, q=2$ and the formula \reff{n03}, one can show
that the last integral is equal to
\begin{equation}\label{n05}
I^{(0)}(\lambda)=a(\lambda).
\end{equation}

Next from the recurrence relations \reff{triv03} we obtain
\begin{equation}\label{n06}
 U_{k}^2={\lambda}^{2}U_{k-1}^2-2\lambda U_{k-1}U_{k-2}+U_{k-2}^2,\quad k\geq 1.
\end{equation}
Substituting in \reff{n02} the right hand side of equality \reff{n06} instead of
$U_{k}^2$ we have ($k\geq 2$)
\begin{equation*}
I^{(k)}(\lambda)=\int_{-2}^{+2}\frac{t^{2}}{(t-\lambda)}U_{k-1}^2\, \mu_{0}^{\prime}(t)\,dt-
2\int_{-2}^{+2}\frac{t}{(t-\lambda)}U_{k-1}U_{k-2}\, \mu_{0}^{\prime}(t)\,dt+I^{(k-2)}(\lambda).
\end{equation*}
Using (\ref{triv405}) we replace  the product  $-tU_{k-1}U_{k-2}$ in this formula by the
expression
$$
1-U_{k-1}^2-U_{k-2}^2,
$$
and obtain
\begin{equation*}
I^{(k)}=\int_{-2}^{+2}\frac{t^{2}}{(t-\lambda)}U_{k-1}^2 \,\mu_{0}^{\prime}(t)\,dt+
2a(\lambda)-2I^{(k-1)}-I^{(k-2)},\quad k\geq 2.
\end{equation*}
From orthonormality of Chebyshev polynomials with respect to $d\mu_{0}$ and that the function 
$tU_{k-1}^{2}(t)\mu_{0}^{\prime}(t)$ is odd it follows that 
the first integral in the right hand side of this equality is equal to
\begin{gather*}
\int_{-2}^{+2}tU_{k-1}^2 \,\mu_{0}^{\prime}(t)\,dt+\lambda\int_{-2}^{+2}U_{k-1}^2 \,\mu_{0}^{\prime}(t)\,dt+
\int_{-2}^{+2}\frac{\lambda^{2}}{(t-\lambda)}U_{k-1}^2 \,\mu_{0}^{\prime}(t)\,dt=\nonumber\\
0+\lambda+\lambda^{2}I^{(k-1)}=\lambda + \lambda^{2}I^{(k-1)}.
\end{gather*}
Then we have
\begin{equation}\label{n07}
I^{(k)}=\lambda+2a+(\lambda^2 -2)I^{(k-1)}-I^{(k-2)},\quad k\geq 2.
\end{equation}
From \reff{n04} we get
\begin{equation*}
\lambda+2a=\frac{a^{2}-1}{a},\quad \lambda^2 -2=\frac{a^{4}+1}{a^2}.
\end{equation*}
Using the relations we can rewrite \reff{n07} in the form
\begin{equation}\label{n08}
I^{(k)}=\frac{a^{2}-1}{a}+\frac{a^{4}+1}{a^2}I^{(k-1)}-I^{(k-2)},\quad k\geq 2.
\end{equation}
Note that when $ k=1 $ the equality \reff{n07} gives
\begin{equation}\label{n09}
I^{(1)}=\lambda+\lambda^{2}I^{(0)}=\lambda+\lambda^{2}a.
\end{equation}

From relations \reff{n05} and \reff{n09} follows the validity of the formula \reff{n02} (and also \reff{n01}) for $k=0$ and  $k=1$. 
Now let us prove that by induction the formula \reff{n02}
is true for any $n\geq 2$.
 We  suppose that \reff{n02} is true for $k=n-1$ and $ k=n-2$ and prove the validity of this formula  for $k=n$.
To do this, we substitute into the right side of equality \reff{n08} with  $k=n$
expression of $I^{(n-1)}$ and $I^{(n-2)}$ obtained from \reff{n02}
at $k=n-1  $ and $ k=n-2$, respectively, and check that it coincides with the right hand side of
\reff{n02} for $k=n$.

We have
\begin{gather}
I^{(n)}=\frac{a^{2}-1}{a}+\frac{a^{4}+1}{a^2}(a+a^3+...+a^{2n-1})-
(a+a^3+...+a^{2n-3})=\nn\\
-\frac{1}{a}+(a^3+...+a^{2n+1})+
(\frac{1}{a}+a+...+a^{2n-3})-(a^3+a^5+...+a^{2n-3})=\nn\\
(a+a^3+...+a^{2n+1})=a\frac{1-a^{2(k+1)}}{1-a^2},\qquad n\geq 2,
\end{gather}
i.e. formula \reff{n02} are valid for $ k=n.$

Now let us prove the validity of formula \reff{n01} for $ n\geq 2$.
Let us suppose that \reff{n01} is true for $k=n-1$ and $ k=n-2.$
Therefore, the right-hand sides of equations \reff{n01} and \reff{n02} coincide when
$k=n-1$ and $k=n-2,$  respectively.
We will show that then the right part of \reff{n01} equal to
the right side \reff{n02} when $k\geq 2$.

Using the formula \reff{n04}, we can rewrite recurrence relations \reff{triv03} in the following form
\begin{equation*}
aU_n=-((a^{2}+1)U_{n-1}+aU_{n-2}),\quad n\geq 2.
\end{equation*}
This allows us to rewrite \reff{n01} as follows
\begin{gather*}
I^{(n)}=(-1)^{n}a^{n+1}U_{n}=(-1)^{n}a^{n}(-(a^{2}+1)U_{n-1}-aU_{n-2})=\\
(-1)^{n-1}a^{n}U_{n-1}(a^{2}+1)-(-1)^{n-2}a^{n-1}U_{n-2}a^{2}=\\
(a^{2}+1)I^{(n-1)}-a^{2}I^{(n-2)}.
\end{gather*}
Substituting the expressions $I^{(n-1)}$ and  $I^{(n-2)}$ using the formula \reff{n02}, we obtain
\begin{gather*}
I^{(n)}=(a^{2}+1)(a+a^3+...+a^{2n-1})-a^{2}(a+a^3+...+
a^{2n-3})=\\
(a^3+...+a^{2n+1})+(a+a^3+...+a^{2n-1})-
(a^3+a^5+...+a^{2n-1})=\\
(a+a^3+...+a^{2n+1})=a\frac{1-a^{2(k+1)}}{1-a^2},\quad n\geq 2.
\end{gather*}
This expression equals to the right hand side of \reff{n02} when $n\geq 2$.
Therefore, the formula \reff{n01} is true for $k=n$ .

\section*{Appendix 2. The proof  of  equation \reff{triv30} when $\vert\lambda\vert \leq 2$}

It is easy to  see that when $\vert\lambda\vert \leq 2$  equality (\ref{triv505}) can be rewritten as
\begin{gather*}
I^{(k)}(\lambda)=v.p.\int_{-2}^{+2}\frac{1}{(t-\lambda)}
\left( U_{k}(\ts\frac{t}{2})\right)^2 \mu_{0}^{\prime}(t)\,dt=
-\frac{\lambda}{2}\left(U_{k}(\ts\frac{\lambda}{2})\right)^2+L_k,
\end{gather*}
where
\begin{equation}\label{triv0012}
L_k=\int_{-2}^{+2}\frac{1}{(t-\lambda)}
\left( U_{k}^{2}(\ts\frac{t}{2})-U_{k}^{2}(\ts\frac{\lambda}{2})\right)\mu_{0}^{\prime}(t)\,dt.
\end{equation}
We want to prove by induction the following formula
\begin{equation}\label{triv0015}
L_{k}(\lambda)=U_{k}(\ts\frac{\lambda}{2})U_{k-1}(\ts\frac{\lambda}{2}),\quad k\geq 0.
\end{equation}
It is obvious that $ L_0=0.$ We compute $ L_1$:
\begin{multline*}
L_1=\!\int_{-2}^{+2}\!\frac{1}{(t-\lambda)}\left( U_{1}^{2}(\ts\frac{t}{2})-
U_{1}^{2}(\ts\frac{\lambda}{2})\right) \mu_{0}^{\prime}(t)\,dt=
\!\int_{-2}^{+2}t\mu_{0}^{\prime}(t)\,dt+\lambda\!\int_{-2}^{+2}\mu_{0}^{\prime}(t)\,dt=
\lambda=U_{1}(\ts\frac{\lambda}{2})U_{0}(\ts\frac{\lambda}{2}),
\end{multline*}
where the last equality follows from \reff{triv03}.

Using formulas \reff{n06} and \reff{triv405},
we rewrite $U_{k}^{2}(\frac{t}{2})-U_{k}^{2}(\frac{\lambda}{2})$ in the following form
\begin{gather}
U_{k}^{2}(\ts\frac{t}{2})-U_{k}^{2}(\ts\frac{\lambda}{2})=
\left(t^{2}U_{k}^{2}(\ts\frac{t}{2})-\lambda^{2}U_{k}^{2}(\ts\frac{\lambda}{2})\right)-\nn\\
\left(2t(-1+U_{k-1}^{2}(\ts\frac{t}{2})+U_{k-2}^{2}(\ts\frac{t}{2}))-
2\lambda U_{k-1}(\ts\frac{\lambda}{2})U_{k-2}(\ts\frac{\lambda}{2})\right)+\nn\\
U_{k-2}^{2}(\ts\frac{t}{2})-U_{k-2}^{2}(\ts\frac{\lambda}{2}).\label{n11}
\end{gather}
Substituting \reff{n11} into (\ref{triv0012}) we get
\begin{equation}\label{triv0013}
L_k=\int_{-2}^{+2}\frac{t^{2}U_{k-1}^{2}(\frac{t}{2})-
\lambda^{2}U_{k-1}^{2}(\frac{\lambda}{2})}{t-\lambda}\mu_{0}^{\prime}(t)\,dt-
2L_{k-1}-L_{k-2}.
\end{equation}
We write $t^{2}U_{k-1}^{2}(\frac{t}{2})-\lambda^{2}U_{k-1}^{2}(\frac{\lambda}{2})$ in the form
\begin{equation}\label{triv0014}
t^{2}U_{k-1}^{2}(\ts\frac{t}{2})-\lambda^{2}U_{k-1}^{2}(\ts\frac{\lambda}{2})=
(t^{2}U_{k-1}^{2}(\ts\frac{t}{2})-
\lambda^{2}U_{k-1}^{2}(\ts\frac{t}{2}))+
(\lambda^{2}U_{k-1}^{2}(\ts\frac{t}{2})-\lambda^{2}U_{k-1}^{2}(\ts\frac{\lambda}{2})).
\end{equation}
Substituting (\ref{triv0014}) into (\ref{triv0013}) we have
\begin{equation}\label{triv0016}
L_k=\lambda+(\lambda^{2}-2)L_{k-1}-L_{k-2},\quad k\geq 2.
\end{equation}

Above we show that \reff{triv0015} is hold for $k=0$ and $k=1.$
Now we prove it by induction for all $k$. For this end we assume that \reff{triv0015}
true for $k=n-1$ and  $k=n-2$ and show that it remains true for $k=n$.

From (\ref{triv0015}) and (\ref{triv0016}) we have
\begin{equation*}
L_k=\lambda+(\lambda^{2}-2)U_{k-1}(\ts\frac{\lambda}{2})U_{k-2}(\ts\frac{\lambda}{2})-
U_{k-2}(\ts\frac{\lambda}{2})U_{k-3}(\ts\frac{\lambda}{2}).
\end{equation*}
Then for the proof of formula (\ref{triv0015}) at $k=n$ it is sufficient to check the validity of
the following equality
\begin{equation*}
\lambda+(\lambda^{2}-2)U_{k-1}(\ts\frac{\lambda}{2})U_{k-2}(\ts\frac{\lambda}{2})-
U_{k-2}(\ts\frac{\lambda}{2})U_{k-3}(\ts\frac{\lambda}{2})=U_{k}(\ts\frac{\lambda}{2})U_{k-1}(\ts\frac{\lambda}{2}).
\end{equation*}
Using recurrent relations (\ref{triv03}), we rewrite the previous equality in the form
\begin{equation*}
\lambda U_{k-1}^{2}(\ts\frac{\lambda}{2})-U_{k-1}(\frac{\ts\lambda}{2})U_{k-2}(\ts\frac{\lambda}{2})=\lambda+
(\lambda^{2}-2)U_{k-1}(\ts\frac{\lambda}{2})U_{k-2}(\ts\frac{\lambda}{2})-
U_{k-2}(\ts\frac{\lambda}{2})U_{k-3}(\ts\frac{\lambda}{2}).
\end{equation*}
It is easy to see that the equality is equivalent to the following equality
\begin{equation*}
\lambda U_{k-1}^{2}=\lambda+\lambda^{2}U_{k-1}U_{k-2}-\lambda U_{k-2}^{2}.
\end{equation*}
Finally, the last equality is equivalent to the identity (\ref{triv405}). Thus
\reff{triv0015} as well as \reff{triv30} are proved.

\section*{Appendix 3. The proof of the identity (\ref{triv405})}

We want to prove by induction the following identity
\begin{equation*}
U_{k}^{2}\big(\ts{\frac{\lambda}{2}}\big)+U_{k-1}^{2}\big(\frac{\lambda}{2}\big)-
\lambda U_{k-1}\big(\frac{\lambda}{2}\big)U_{k}\big(\frac{\lambda}{2}\big)=1,
\end{equation*}
It is obvious that the equality holds for $k=0$.
 We will show that for any $n\geq 1 $ this equation is valid when
$k=n$, if it is true for $k=n-1$. Using recurrent relation (\ref{triv03}), we have
\begin{gather*}
U_{n}^{2}+U_{n-1}^{2}-\lambda U_{n-1}U_{n}=\\
(\lambda U_{n-1}-U_{n-2})^{2}+U_{n-1}^{2}-\lambda U_{n-1}(\lambda U_{n-1}-U_{n-2})=\\
U_{n-1}^{2}+U_{n-2}^{2}-\lambda U_{n-1}U_{n-2}.
\end{gather*}
The obtained expression is equal to unity according to the inductive assumption.
Hence, the formula (\ref{triv405}) is proved.

\section*{Appendix 4. The proof of the formula \reff{t25}}
Here we construct the polynomials $\{\phi_{n}^{(k)}(\lambda)\}_{n=0}^{\infty}$ related to
the Jacobi matrix $\mathcal{J}_{k}^{(\beta)}$. These polynomials (see \reff{triv1}, \reff{triv1`})
is determined by following recurrence relations ($n\geq 0$)
\begin{gather}
\lambda\phi_{n}^{(k)}(\ts\frac{\lambda}{2})=
\phi_{n+1}^{(k)}(\ts\frac{\lambda}{2})+a_{n}\phi_{n}^{(k)}(\ts\frac{\lambda}{2})+
b_{n-1}\phi_{n-1}^{(k)}(\ts\frac{\lambda}{2}),\label{triv111} \\
\,\phi_{0}^{(k)}=1,\quad a_n =\beta\delta_{n k},\quad b_n =1- \delta_{-1 n}.\nn
\end{gather}
From recurrent relations (\ref{triv03}) and (\ref{triv111}), 
it follows that
\begin{equation}\label{triv112}
\phi_{0}^{(k)}=U_{0},\quad \phi_{1}^{(k)}(\ts\frac{\lambda}{2})=U_{1}(\ts\frac{\lambda}{2}),\,\ldots\,,\,
\phi_{k}^{(k)}(\ts\frac{\lambda}{2})=U_{k}(\ts\frac{\lambda}{2}).
\end{equation}
From (\ref{triv112}), using the recurrent relations (\ref{triv03}) and \ref{triv111}
one obtains by induction that
\begin{gather*}
\phi_{k+1}^{(k)}=U_{k+1}-\beta U_{k},\\
\phi_{k+2}^{(k)}=U_{k+2}-\beta U_{k+1}-\beta U_{k-1},...,\\
\phi_{2k+1}^{(k)}=U_{2k+1}-\beta U_{2k}-\beta U_{2k-2}-...-\beta U_{0}.
\end{gather*}
Next, we have the following relation
\begin{gather}
\phi_{2k+2}^{(k)}(\ts\frac{\lambda}{2})=\lambda\phi_{2k+1}^{(k)}(\ts\frac{\lambda}{2})-
\phi_{2k}^{(k)}(\ts\frac{\lambda}{2})=\nn\\
\lambda(U_{2k+1}-\beta U_{2k}-\beta U_{2k-2}-...-\beta U_{0})-
(U_{2k}-\beta U_{2k-1}-\beta U_{2k-3}-...-\beta U_{1}).\label{triv114}
\end{gather}
Combining terms in pairs, standing on the same locations in the first and second
brackets in the right hand side of equality \reff{triv114}, and using recurrence relations (\ref{triv03}), we get
\begin{equation*}
\phi_{2k+2}^{(k)}=U_{2k+2}-\beta U_{2k+1}-\beta U_{2k-1}-...-\beta U_{1}.
\end{equation*}
Finally, to check the validity of the formula
\begin{equation}\label{triv115}
\phi_{n}^{(k)}=U_{n}-\beta U_{n-1}-\beta U_{n-3}-...-\beta U_{n-(2k+1)},\qquad
 n\geq 2k+3,\quad k\geq 0\,,
\end{equation}
it is enough to check that the polynomials $\phi_{n}^{(k)}$ defined by equations \reff{triv115}
for $n\geq {2k+3}$ satisfy recurrence relations \reff{triv111} that for $n\geq k$ have the form
\begin{equation*}
\phi_{n}^{(k)}(\ts\frac{\lambda}{2})=
\lambda\phi_{n-1}^{(k)}(\ts\frac{\lambda}{2})-\phi_{n-2}^{(k)}(\ts\frac{\lambda}{2}).
\end{equation*}
Indeed, we have for all $n\geq {2k+3}$, $ k\geq 0 $
\begin{multline*}
\phi_{n}^{(k)}-\lambda\phi_{n-1}^{(k)}+\phi_{n-2}^{(k)}=\\
U_{n}-\beta U_{n-1}-\beta U_{n-3}-...-\beta U_{n-(2k+1)}-
\lambda(U_{n-1}-\beta U_{n-2}-\beta U_{n-4}-...-\beta U_{n-2k-2})+\\
(U_{n-2}-\beta U_{n-3}-\beta U_{n-5}-...-\beta U_{n-2k-3})=\qquad\qquad\\
(U_{n}-\lambda U_{n-1}+ U_{n-2})-\beta(U_{n-1}-\lambda U_{n-2}+ U_{n-3})-\ldots\qquad\qquad\\
\ldots-\beta(U_{n-(2k+1)}-\lambda U_{n-2(k+1)}+ U_{n-2k-3})=0.
\end{multline*}
Thus, we proved that for all $k\geq 0 $, the polynomials $\phi_{n}^{(k)}$ can be calculated
by the following formulas
\begin{gather*}
\phi_{0}^{(k)}=U_{0},\quad \phi_{1}^{(k)}=U_{1},\ldots,\phi_{k}^{(k)}=U_{k},\\
\phi_{k+s}^{(k)}=U_{k+s}-\beta U_{k+s-1}-\beta U_{k+s-3}-\ldots-\beta U_{k-s+1},
\qquad 1\leq s\leq k+1,\\
\phi_{n}^{(k)}=U_{n}-\beta U_{n-1}-\beta U_{n-3}-...-\beta U_{n-(2k+1)},
\qquad n\geq 2k+2.
\end{gather*}

\end{document}